\documentclass{svjour2}
\usepackage{graphicx}
\usepackage{latexsym}
\usepackage{amssymb}

\def\beq{\begin{equation}}
\def\eeq{\end{equation}}
\def\eqref#1{(\ref{#1})}

\journalname{General Relativity and Gravitation}

\begin{document}

\title{Scattering of uncharged particles in the field of two extremely charged black holes}

\author{Donato  Bini \and Andrea Geralico \and Gabriele Gionti \and Wolfango Plastino \and Nelson Velandia
}

\institute{
Donato Bini
\at
Istituto per le Applicazioni del Calcolo ``M. Picone,'' CNR, I--00161 Rome, Italy\\
INFN, Sezione di Roma Tre, I-00146 Rome, Italy
\and
Andrea Geralico 
\at
Istituto per le Applicazioni del Calcolo ``M. Picone,'' CNR, I--00161 Rome, Italy
\and
Gabriele Gionti, S.J. 
\at
Specola Vaticana, V-00120 Vatican City, Vatican City State\\
Vatican Observatory Research Group Steward Observatory, The University Of Arizona, 
933 North Cherry Avenue, Tucson, Arizona 85721, USA\\
INFN, Laboratori Nazionali di Frascati, Via E. Fermi 40, 00044 Frascati, Italy
\and
Wolfango Plastino 
\at
Roma Tre University, Department of Mathematics and Physics, I-00146 Rome, Italy\\
INFN, Sezione di Roma Tre, I-00146 Rome, Italy
\and
Nelson Velandia
\at
Departamento de Fisica, Facultad de Ciencias Pontificia Universidad Javeriana, 110231
Bogot\'a D. C., Colombia
}

\date{Received: date / Accepted: date / Version: \today}

\maketitle

\begin{abstract}
We investigate the motion of uncharged particles scattered by a binary system consisting of extremely charged black holes in equilibrium as described by the Majumdar-Papapetrou solution. 
We focus on unbound orbits confined to the plane containing both black holes.
We consider the two complementary situations of particles approaching the system along a direction parallel to the axis where the black holes are displaced and orthogonal to it.
We numerically compute the scattering angle as a function of the particle's conserved energy parameter, which provides a gauge-invariant information of the scattering process.
We also study the precession of a test gyroscope along such orbits and evaluate the accumulated precession angle after a full scattering, which is another gauge-invariant quantity. 
\end{abstract}

\section{Introduction}

The interaction between self-gravitating compact objects is very complicated and difficult to be investigated in its complete generality by using either analytical and numerical methods, so that one can only be content with special cases or approximated situations. 
The dynamics of a two-body system (spinless or with spin, neutral or charged, even with additional multipolar structure) has been largely studied within both the Post-Newtonian (PN) \cite{Bini:2017wfr} and Post-Minkowskian (PM) \cite{Bel:1981be,Westpfahl:1985} approximation of general relativity, with also some post-Schwarzschild treatment \cite{Damour:2016gwp,Bini:2017xzy,Bini:2018ywr,Bern:2019nnu,Antonelli:2019ytb}.
In the first case, PN,  one has reached the 4PN-order level of approximation in the weak-field and slow-motion limit, overcoming a lot of related technical difficulties. In the second case, PM,  one has reached the 2PM-order level of approximation in the weak-field limit. 

Other common approximation tools are both metric and curvature perturbations due to a small body in the exact gravitational field of the larger one (a Schwarzschild or a Kerr black hole), if the mass of one body is much smaller than the other. This is the approximation adopted by the Gravitational Self-Force (GSF) theory (see e.g., Ref. \cite{Detweiler:2005kq}). In spite of the big simplification with respect to the original general problem, this approach has its own difficulties too. For example, renormalization techniques should be used to avoid field singularities at the location of the perturbing body.
A lesson we have learned up to now is that all possible information concerning the physics of binary systems, whatever is the context in which each of them has been obtained, should be combined together not losing even a bit of that information and using  them constructively as much as possible. In this direction moves, for example, the Effective-One-Body (EOB) formalism, started about 20 years ago by Buonanno and Damour (see, e.g., the pioneering and seminal papers \cite{Buonanno:1998gg,Buonanno:2000ef}). Similarly, one has other approaches to study the physics of the system, including effective field theories (see, e.g., Ref. \cite{Foffa:2013qca} and references therein), scattering amplitudes (see, e.g., \cite{Bern:2017yxu,Bern:2019nnu} and references therein), etc.

The first motivation for this study is given by the recent detection of gravitational wave signals from a binary system of two black holes as well as two neutron stars. Such signals are expected to come 1) from coalescing phenomena: one body under the attraction of the other starts spiraling around it emitting gravitational energy and the collapsing on it giving rise to a single object (a black hole mainly) with a certain mass and spin; 2) hyperbolic encounters or scattering phenomena: the second body is energetic enough to avoid the capture by the first one, but at the minimum approach distance (mainly)  it emits a lot of radiation which can hopefully be detected too.

The complete knowledge of the conservative dynamics in the test field approximation is preliminary to the study of backreaction effects due to gravitational wave emission within the GSF framework. 
The simplest problem of test particles moving along hyperbolic-like geodesic orbits in black hole spacetimes is fully known. 
In previous works we have studied the scattering of test particles endowed with spin or undergoing self-interaction effects by Schwarzschild and Kerr black holes \cite{Bini:2016tqz,Bini:2016ubc,Bini:2017ldh,Bini:2017pee,Bini:2018zxp}.
In such cases  nongeodesic motion has been considered, and the deviation from the geodesic orbits due to additional structure or different kind of interaction was encoded in the scattering angle, i.e., the most natural gauge-invariant and physical observable associated with the scattering process, as a (small) correction to the geodesic scattering angle.

In the present paper we will investigate the features of hyperbolic-like motion in the spacetime of two (extreme) Reissner-Nordstr\"om black holes belonging to the Majumdar-Papapetrou class of static axially symmetric electrovacuum solutions to the Einstein-Maxwell field equations \cite{Hartle:1972ya}. The black holes are extremely charged, i.e., their charge-to-mass ratios are both equal to one, and equilibrium exists independent of the separation between the bodies, due to the balance between gravitational attraction and electrostatic repulsion. 
Therefore, such a solution allows one to analytically explore the limiting situation when the two bodies are very close each other.  
In addition, this static \lq\lq equilibrium" configuration may be considered as an instantaneous snapshot of the actual head-on collision of two black holes, at least in a regime where the two bodies are well separated and approach each other with velocities much smaller than the speed of light, and may help to better understand some features of the corresponding dynamical situation when the two black holes interact (also in absence of charges).

The Majumdar-Papapetrou solution has given a renewed interest in the last few years.
Particular attention has been devoted to bound null geodesic orbits, in order to investigate the qualitative features of binary black hole shadows \cite{Shipley:2016omi} as well as the existence of photon surfaces and closed photon orbits around such binaries, leading, e.g., to strong lensing effects \cite{Assumpcao:2018bka}.
Bound timelike orbits have been studied, e.g., in Refs. \cite{Wunsch:2013st,Ryzner:2015jda}.
Here we will consider instead those unbound timelike orbits of  particles starting far from the binary, approaching it up to a minimum approach distance, and then escaping undergoing a scattering process, the motion being confined on a plane containing both black holes.  
Since we are mainly interested on purely gravitational effects, we will consider neutral test particles in order to neglect electromagnetic interactions.
We will select two different families of orbits, corresponding to the two complementary situations of particles approaching the system along a direction parallel to the axis where the black holes are displaced and orthogonal to it.
We will numerically compute two gauge-invariant quantities characterizing the scattering process: the scattering angle as well as the accumulated precession angle of a test gyroscope along such orbits after a full scattering.
In spite of the apparent simplicity of the metric, the geodesics (and then the orbits of charged particles too) cannot be separated, even in the null case which usually implies additional simplifications with respect to the corresponding timelike case.
This feature complicates matters, and the consequence is that one is forced to perform most of the analysis numerically.

The metric signature here is mostly positive, $-+++$. Greek indices run from 0 to 3, whereas Latin indices from 1 to 3.
We will use geometric units such that $c=1=G$.

\section{The Majumdar-Papapetrou solution}

The  Majumdar-Papapetrou solution for a system of two (extreme) Reissner-Nordstr\"om black holes with masses $M_1$, $M_2$ and charges $|Q_1|=M_1$, $|Q_2|=M_2$ in Cartesian-like coordinates $x^\alpha =\{t,x,y,z\}$   reads
\beq
\label{MP_2bh}
ds^2=-\frac{dt^2}{U^2}+U^2\delta_{ab}dx^adx^b\,,
\eeq
 with $x^a=\{x,y,z\}$, a=1,2,3 and with a vector potential $A$ and electromagnetic tensor $F=dA$ given by
\beq
A=U^{-1}dt\,,\qquad F=dA=-\frac{U_{,\alpha}}{U^2}dt\wedge dx^\alpha\,.
\eeq
Here 
\beq
U=1+\frac{M_1}{r_1}+\frac{M_2}{r_2}\,,
\eeq
with 
\beq
r_1=\sqrt{x^2+y^2+\left(z-\frac{b}2\right)^2}\,,\qquad
r_2=\sqrt{x^2+y^2+\left(z+\frac{b}2\right)^2}\,.
\eeq
The black holes are located along the $z$-axis at a distance $b$ from each other and displaced symmetrically with respect to the $z=0$ hyperplane, which then becomes a symmetry plane for the whole metric in the equal mass case.
In these coordinates the event horizon thus consists of two disconnected components lying at the spatial points $(0,0,\pm\frac{b}2)$. The properties of some associated curvature as well as electromagnetic invariants have been recently studied in Ref. \cite{Semerak:2016gfz}.

Let us choose as a fiducial observer family the static observers, with 4-velocity $u\equiv e_{\hat 0}$ ($e_{\hat 0}^\flat$ denoting its fully covariant form)
\beq
e_{\hat 0}\equiv u = U\partial_t\,,\qquad e_{\hat 0}^\flat = -U^{-1}dt\,,
\eeq
and adapted orthonormal spatial triad 
\beq
\label{thdframe}
e_{\hat a}=U^{-1}\partial_a\,, \qquad e_{\hat a}\cdot e_{\hat b} =\delta_{ab},\quad a=x,y,z\,.
\eeq
The dual frame $\omega^{\hat \alpha}$ such that $\omega^{\hat \alpha}(e_{\hat \beta})=\delta^\alpha_\beta$ is given by
\beq
\omega^{\hat 0}=U^{-1}dt\,,\qquad \omega^{\hat a}=Udx^a\,.
\eeq
The observers $u$ are accelerated, with acceleration 
\beq
a(u)^\flat=-(\partial_\alpha \ln U ) \, dx^\alpha\,,
\eeq
whereas their vorticity and expansion vanish identically \cite{Jantzen:1992rg}.
When measured by the observer $u$ the electromagnetic field is a purely electric field
\beq
F=u^\flat \wedge E(u)^\flat \,,\qquad E(u)^\flat=(\partial_\alpha \ln U ) \, dx^\alpha\,,
\eeq
and the electric field is the negative of the observer's acceleration, so that
\beq
F=-u^\flat \wedge a(u)^\flat\,.
\eeq
Using the gravito-electromagnetic terminology the electric field coincides with the gravito-electric one \cite{Jantzen:1992rg}, i.e., $E(u)=g(u)$, since 
$g(u)=-a(u)$, implying that electric and gravito-electric field with respect to $u$ coincide.
Similarly, the Riemann tensor is summarized by its electric part ${\mathcal E}(u)_{\alpha\gamma}=R_{\alpha\beta\gamma\delta}u^\beta u^\delta$ only, since the magnetic one vanishes identically.
Furthermore, ${\mathcal E}(u)$ satisfies the symmetric curl property ${\rm Scurl}(U^{-2}{\mathcal E}(u))=0$, typical of static spacetimes (see Ref. \cite{Bini:2004qf} for notational details).

\subsection{Curvature invariants}

A natural Newman-Penrose frame is built as follows
\beq
l=\frac{1}{\sqrt{2}}(e_{\hat 0}+e_{\hat x})\,,\quad n=\frac{1}{\sqrt{2}}(e_{\hat 0}-e_{\hat x})\,,\quad
m=\frac{1}{\sqrt{2}}(e_{\hat y}+i e_{\hat z})\,.
\eeq
The corresponding Weyl scalars are given by
\begin{eqnarray}
\psi_0 &=&\bar \psi_4= \frac12(V_{zz}-V_{yy})-iV_{yz}\,,\nonumber\\
\psi_1 &=&-\bar \psi_3=  \frac12(V_{xy}+iV_{xz})\,,\nonumber\\
\psi_2 &=& -\frac16 (2V_{xx}-V_{yy}-V_{zz}) \,,
\end{eqnarray}
where we have used the notation
\beq
V_{ab}=-\frac12\partial_{ab}(U^{-2})\,.
\eeq
With the $\psi_j$ ($j=0\ldots 4$) one forms the two invariants
\begin{eqnarray}
I&=&|\psi_0|^2+4|\psi_1|^2+3\psi_2^2
\,,\nonumber\\
J&=&{\rm det}\,\,
\pmatrix{ 
 \psi_4 &\psi_3&\psi_2\cr
 \psi_3 & \psi_2& \psi_1\cr
 \psi_2 &\psi_1& \psi_0
}\nonumber\\
&=&(|\psi_0|^2-2|\psi_1|^2-\psi_2^2)\psi_2-2{\rm Re}(\psi_0\bar\psi_1^2)
\,,
\end{eqnarray}
and then the so-called speciality index \cite{Baker:2000zm}
\beq
S=\frac{27 J^2}{I^3} \,,
\eeq
which is a (gauge-invariant) indicator of the algebraic characterization of the spacetime in terms of the multiplicity of its \lq\lq principal null directions," i.e., those directions which are the eigendirections of the Weyl tensor itself.
In particular, $S=1$ when the spacetime is algebraically special.

For generic values of the parameters the Majumdar-Papapetrou solution is of general Petrov type I with $S\not =1$.
However, the latter reduces to 1 satisfying the condition for algebraic speciality in both limits of small ($b\to0$) and large separations ($b\to \infty$) (besides the trivial limit of vanishing mass for one of the two bodies).
For later purposes, let us suppress the $y-$dependence focusing on the meridian section $y=0$. 
The relevant (rescaled) Weyl scalars then reduce to
\begin{eqnarray}
\tilde\psi_0 &=&\frac38\left[
(b-2z)^2\frac{M_1}{r_1^5}+(b+2z)^2\frac{M_2}{r_2^5}
+\frac14\frac{M_1}{r_1^5}\frac{M_2}{r_2^5}b^2(4x^2+b^2-4z^2)^2
\right]\,,\nonumber\\
\tilde\psi_1 &=&-\frac34ix\left[
(b-2z)\frac{M_1}{r_1^5}-(b+2z)\frac{M_2}{r_2^5}
+\frac{M_1}{r_1^5}\frac{M_2}{r_2^5}b^2z(4x^2+b^2-4z^2)
\right]\,,\nonumber\\
\tilde\psi_2 &=& \frac13 \tilde\psi_0-x^2\left(
\frac{M_1}{r_1^5}+\frac{M_2}{r_2^5}+4\frac{M_1}{r_1^5}\frac{M_2}{r_2^5}b^2z^2
\right) \,,
\end{eqnarray}
where $\tilde\psi_j=U^4\psi_j$ and the functions $U$ and $r_{1,2}$ are evaluated at $y=0$.
On the symmetry axis ($x=0$) the previous expressions further simplify so that $\psi_1=0$ and $\psi_2=\frac13\psi_0$, implying $S=1$ for every choice of the parameters.

\begin{figure}
\begin{center}
$\begin{array}{cc}
\includegraphics[scale=0.3]{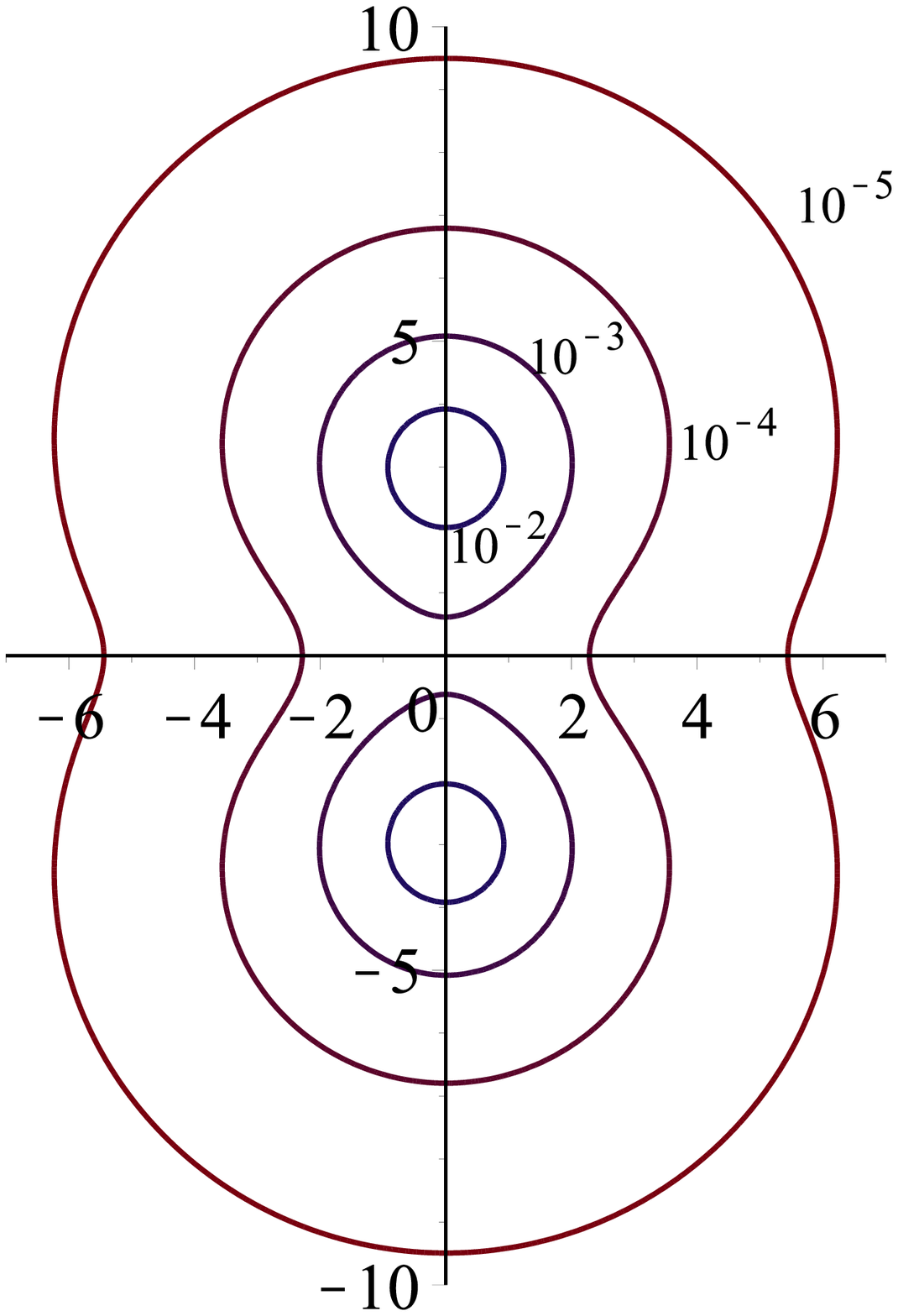}&\qquad
\includegraphics[scale=0.3]{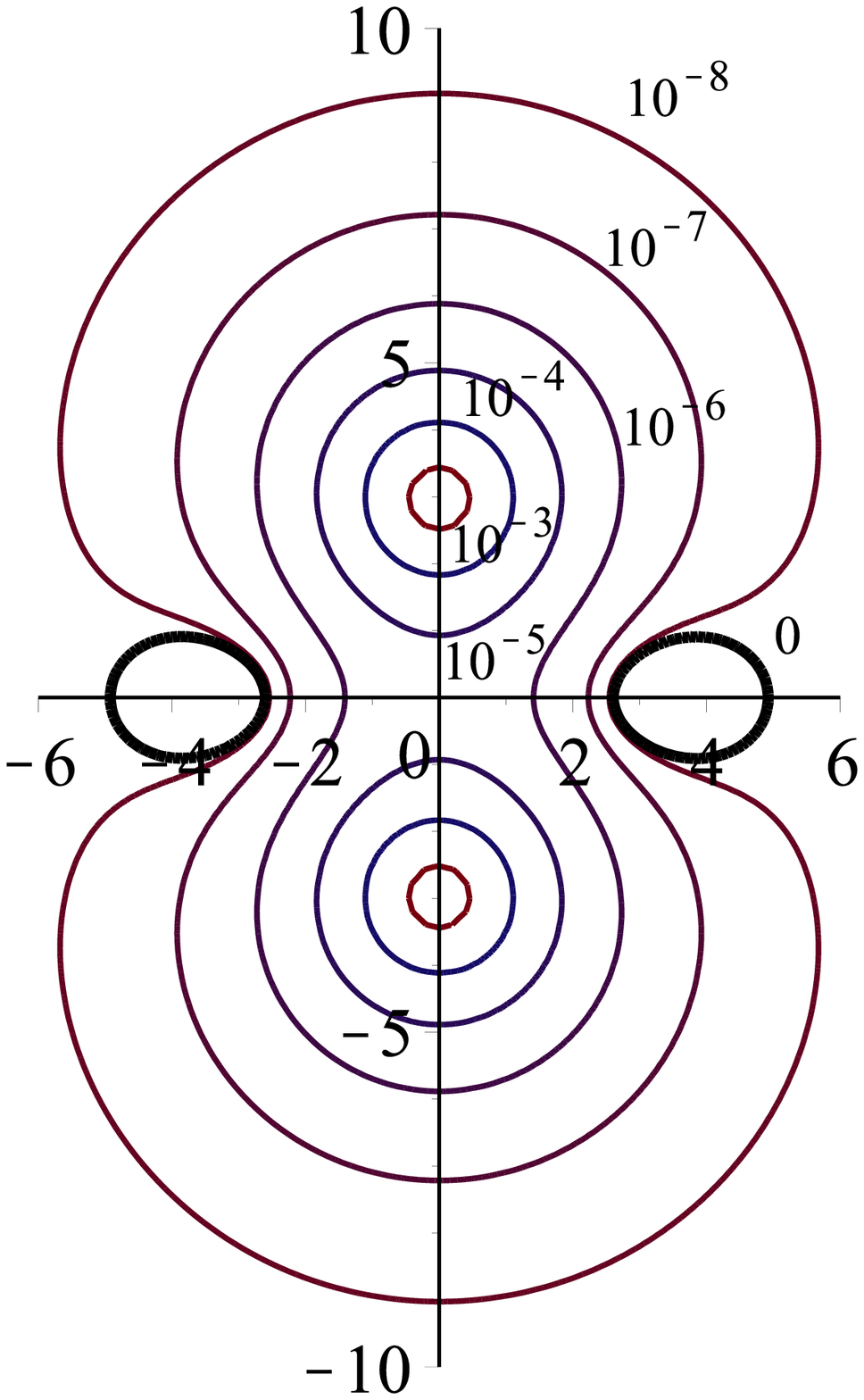}\\[.2cm]
\mbox{(a)} &\qquad \mbox{(b)}\cr
\end{array}
$\\[.5cm]
\includegraphics[scale=0.6]{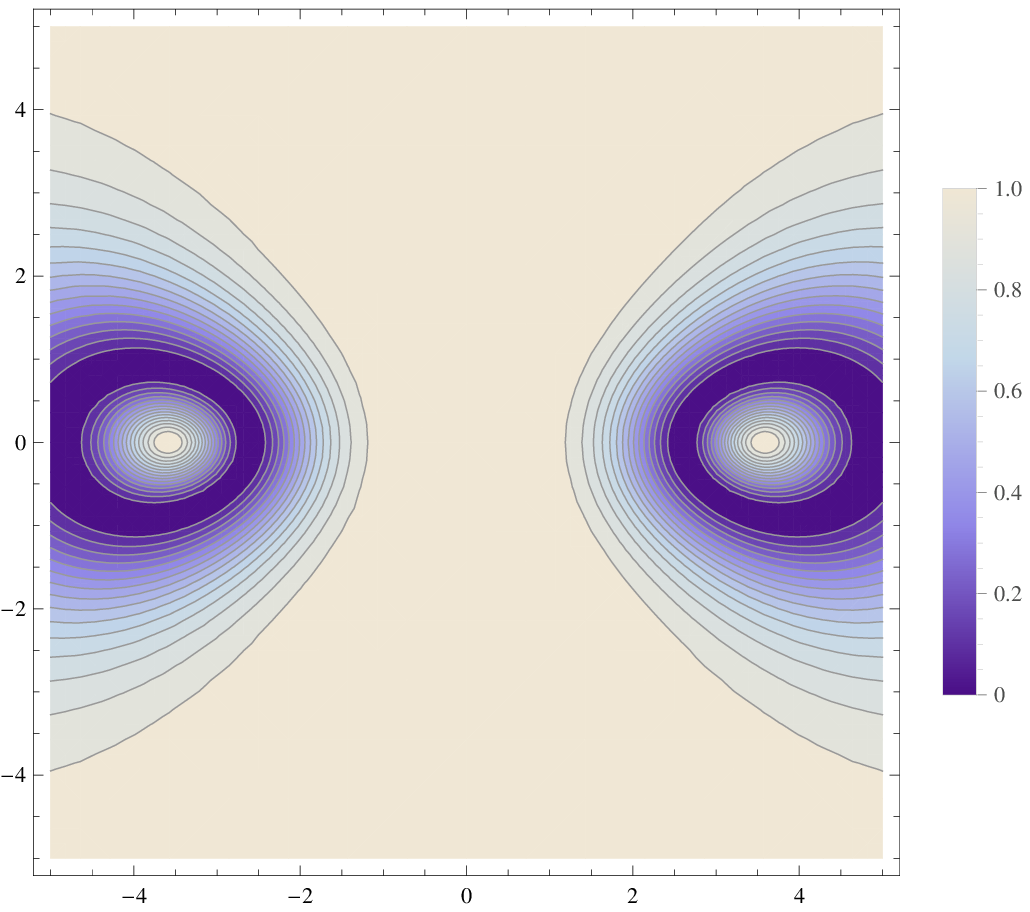}\\[.2cm]
\mbox{(c)}
\end{center}
\caption{The curvature invariants $I$ and $J$ and the speciality index $S\in[0,1]$ on the (meridian) $x-z$ plane are plotted in the case of equal mass black holes $M_1=1=M_2$ at coordinate separation $b=6$ in panels (a) to (c), respectively. 
}
\label{fig:speciality}
\end{figure}

We show in Fig. \ref{fig:speciality} (a) to (c) the contours of the invariants $I$, $J$ and $S$, respectively, on the (meridian) $x-z$ plane containing the two black holes and for fixed values of the parameters.
The value of $S$ is unity almost everywhere except in the neighborhood of the curves where $I$ vanishes.
The latter are located symmetrically with respect to the symmetry axis, and move away from it for increasing values of the separation distance between the black holes.
For equal mass black holes these curves are also symmetric about the $x-$axis, whereas this symmetry is lost in the unequal mass case. Furthermore, in this case they are no more centered on the $x-$axis ($z=0$), but their centers move down to negative values of $z$ for $M_2<M_1$ and up to positive values of $z$ for $M_2>M_1$.


\subsection{Timelike geodesics in the meridian plane}

Geodesic motion in the Majumdar-Papapetrou spacetime was first investigated by Chandrasekhar \cite{Chandrasekhar:1989vk} and Contopoulos \cite{Contopoulos:1990,Contopoulos:1991}. The geodesic equations cannot be separated in general, but they become integrable in the weak field limit \cite{Contopoulos:1993}.  
Further studies have shown that both photon and massive particle motion in the Majumdar-Papapetrou spacetime exhibits self-similarity properties as well as features typical of chaotic dynamics \cite{Dettmann:1994dj,Yurtsever:1994yb,Dettmann:1995ex,Contopoulos:1999js,Alonso:2007ts}.

Let us consider timelike geodesics (associated with neutral massive particles) characterized by the 4-velocity
\beq
U_{\rm (geo)}=\frac{dx^\alpha}{d\tau}\partial_\alpha = U_{\rm (geo)}^\alpha\partial_\alpha\,,
\eeq
with
\beq
U_{\rm (geo)}^t=\frac{dt}{d\tau}=U^2 E\,,
\eeq
where $E$ denotes the conserved Killing energy per unit particle's mass such that $U_{\rm (geo)}{}_t=-E$. 
There exists another constant of motion, i.e., the angular momentum per unit mass along the $z-$axis 
\beq
L=xU_{\rm (geo)}{}_y-yU_{\rm (geo)}{}_x
= U^2\left(x\frac{dy}{d\tau}-y\frac{dx}{d\tau}\right)\,,
\eeq
associated with the Killing vector $\xi=x\partial_y-y\partial_x$.
We are interested in those orbits confined in a plane containing the black holes, i.e., the symmetry axis.
Therefore, we set $U_{\rm (geo)}^y=0$ and $y=0$ without loss of generality, corresponding to the meridian $x-z$ plane where $L=0$.
The normalization condition $U_{\rm (geo)} \cdot U_{\rm (geo)} =-1$ then implies
\beq
(U_{\rm (geo)}^x)^2+(U_{\rm (geo)}^z)^2=E^2-\frac{1}{U^2}\,,
\eeq
and hence the final form for $U_{\rm (geo)}$ is
\beq
U_{\rm (geo)}=E U^2\, \partial_t \pm \left[E^2-\frac{1}{U^2}- (U_{\rm (geo)}^z)^2 \right]^{1/2}\, \partial_x +U_{\rm (geo)}^z \partial_z \,,
\eeq
with $U_{\rm (geo)}^z=\frac{dz}{d\tau}$.

When referring $U_{\rm (geo)}$ to the static observer frame we have
\beq
U_{\rm (geo)}=\gamma [u+\nu^{\hat a}e_{\hat a}]\,,
\eeq
with
\beq
\gamma=EU\,,\qquad \nu^{\hat x}=\pm  \left[1-\frac{1}{E^2 U^2}-\left(\frac{U_{\rm (geo)}^z}{E}\right)^2 \right]^{1/2} \,,\qquad \nu^{\hat z}=   \frac{U_{\rm (geo)}^{z}}{E}\,,
\eeq
so that
\beq
U_{\rm (geo)}^t=\gamma U\,,\qquad U_{\rm (geo)}^x=E\nu^{\hat x}\,,\qquad U_{\rm (geo)}^z=E \nu^{\hat z}\,.
\eeq
The magnitude of the relative (spatial with respect to $u$) velocity $\nu$ is given by
\beq
\nu^2=(\nu^{\hat x})^2+(\nu^{\hat z})^2=1-\frac{1}{E^2 U^2}\,.
\eeq
Equivalently, one can introduce the unit vector of the spatial velocity
\beq
\hat \nu = \cos \alpha e_{\hat x}+\sin \alpha e_{\hat z}\,,\qquad \alpha=\alpha(\tau)\in [0,\pi]\,,
\eeq 
so that
\beq
U_{\rm (geo)}=EU \left[u+\sqrt{1-\frac{1}{E^2 U^2}}\,\, \hat \nu \right]\,.
\eeq
The geodesic equations are conveniently written by using the notation
\beq
g_{\hat x} =\frac{\partial_x U}{U}\,,\qquad g_{\hat z} =\frac{\partial_z U}{U}\,,
\eeq
that is
\begin{eqnarray}
\frac{dx}{d\tau}&=&E\nu\cos\alpha\,,\qquad
\frac{dz}{d\tau}=E\nu\sin\alpha\,,\nonumber\\
\frac{d\alpha}{d\tau}&=&\frac{E(1+\nu^2)}{\nu}\left(-\sin\alpha g_{\hat x}+\cos\alpha g_{\hat z}\right)\,,
\end{eqnarray}
where $\nu$ is a function of $\tau$ 
\beq
\nu=\nu(\tau)=\sqrt{1-\frac{1}{E^2 U(\tau)^2}}\,,\qquad U(\tau)=U|_{x^a=x^a(\tau)}\,,
\eeq
and depends on the particle's energy parameter $E$ and, through $U$, also on the spacetime parameters $M_1$, $M_2$ and $b$. 
This system can be integrated numerically with initial conditions $x(0)=x_0$, $z(0)=z_0$ and $\alpha(0)=\alpha_0$ (back and forth from the initial value $\tau=0$ of the proper time) for fixed values of the parameters.

\section{Scattering angle}

\begin{figure}
\begin{center}
$\begin{array}{cc}
\includegraphics[scale=0.25]{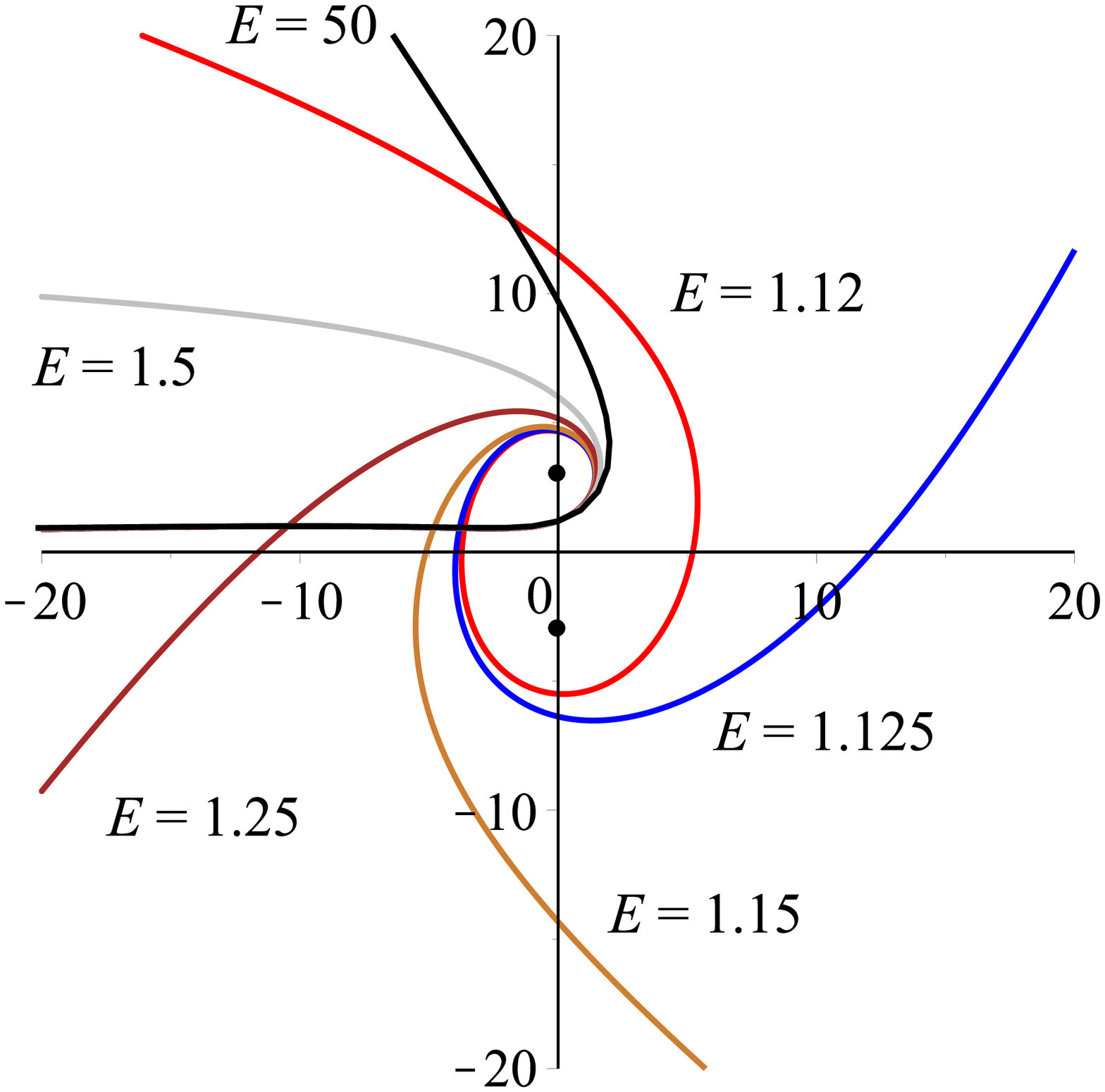}&\qquad
\includegraphics[scale=0.25]{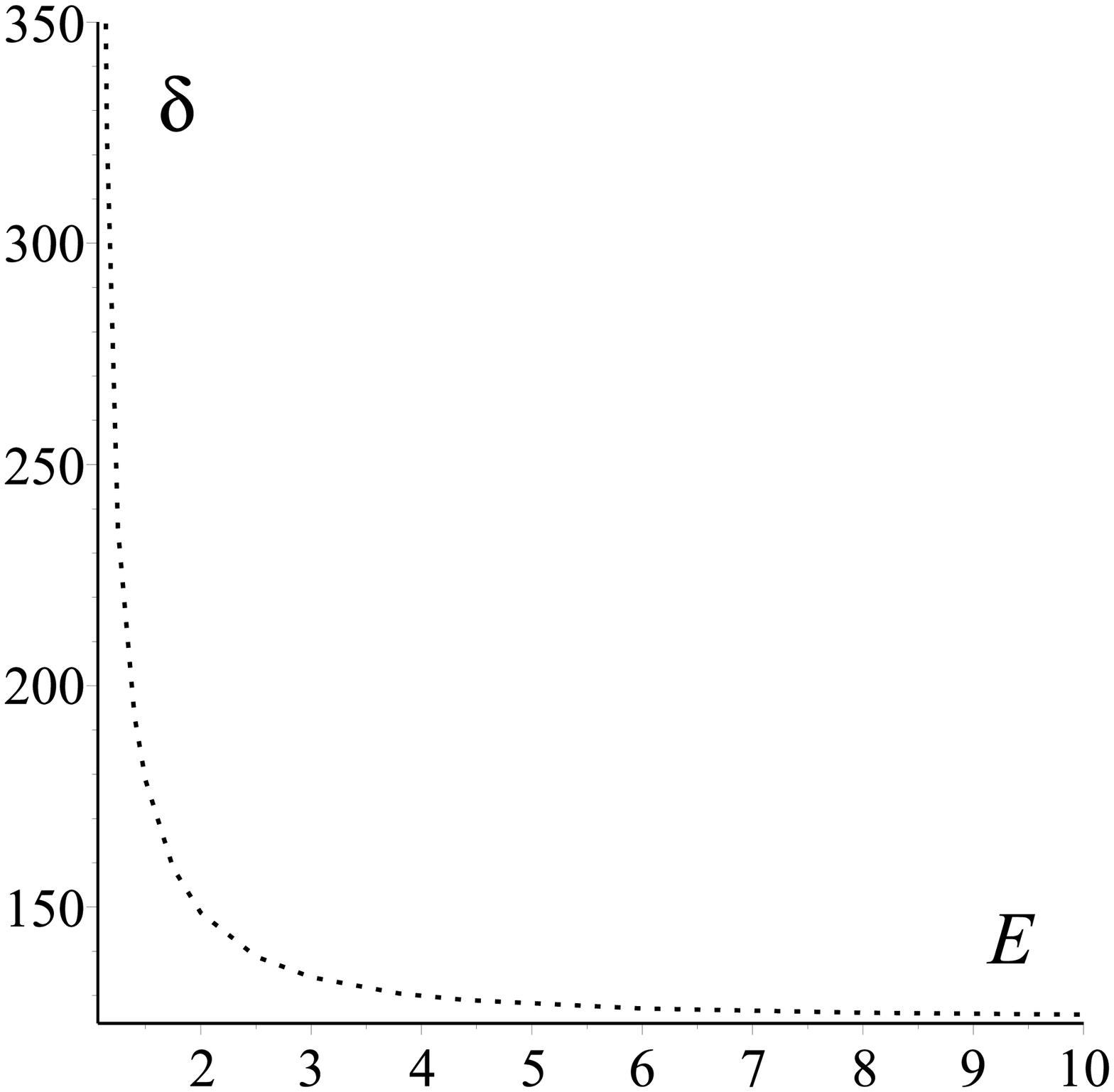}\\[.2cm]
\mbox{(a)} &\qquad \mbox{(b)}\cr
\end{array}
$\\
\end{center}
\caption{
Examples of numerical integration of scattering orbits in the $x-z$ plane are shown in panel (a) with the parameter choice $M_1=M_2=M=1$, $b/M=6$ and initial conditions $x(0)=-10$, $z(0)=1$ and $\alpha(0)=0$ for every fixed value of energy.
The behavior of the scattering angle $\delta$ as a function of the energy is shown in panel (b).
For $M_2<M_1$ the curve moves to the right, since more energy is needed to escape the attraction by the black hole with larger mass $M_1$ (located in the same half plane ($z>0$) as the approaching trajectory).
}
\label{fig:orbite_asse_x}
\end{figure}

\begin{figure}
\begin{center}
$\begin{array}{cc}
\includegraphics[scale=0.25]{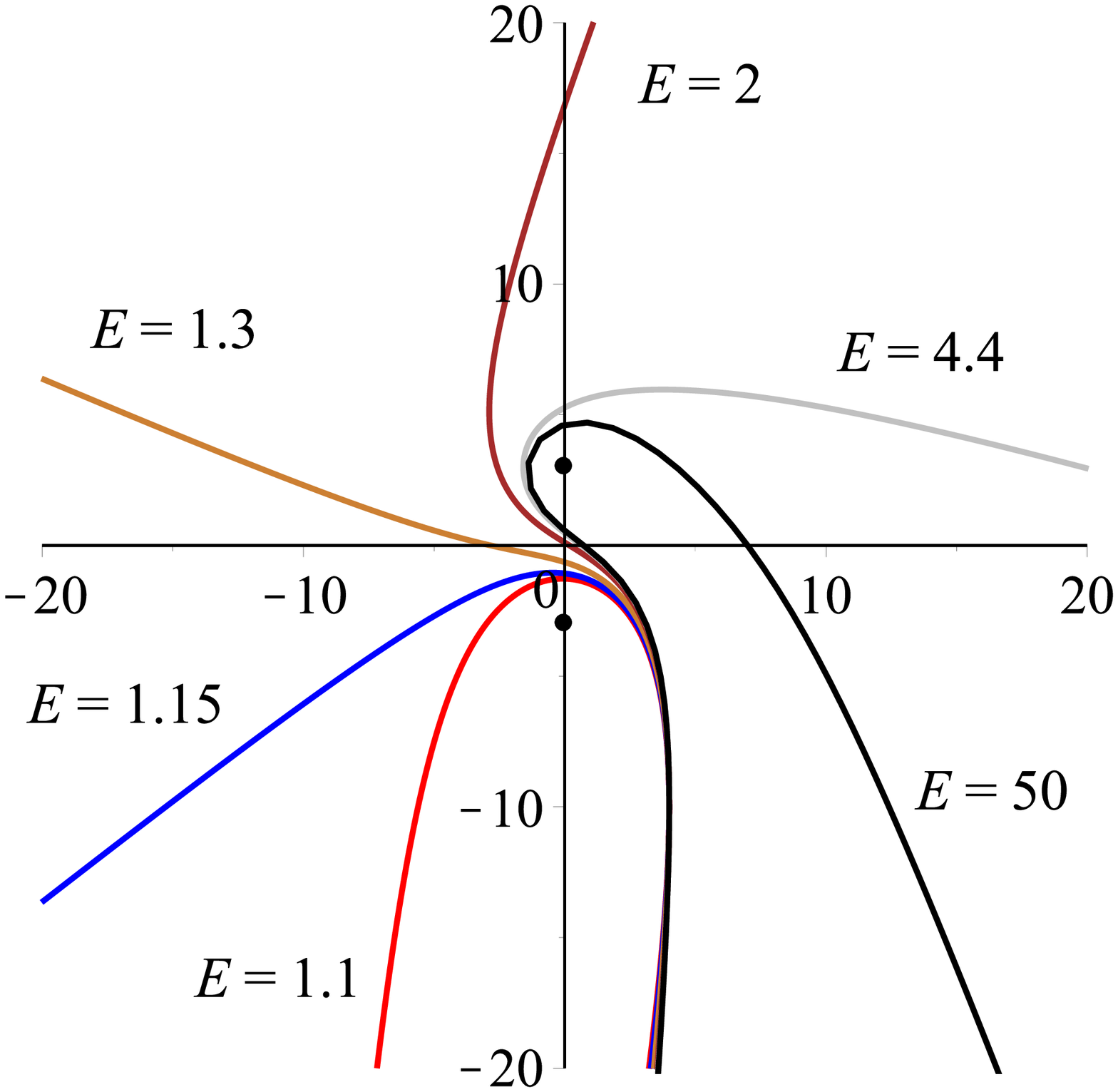}&\qquad
\includegraphics[scale=0.25]{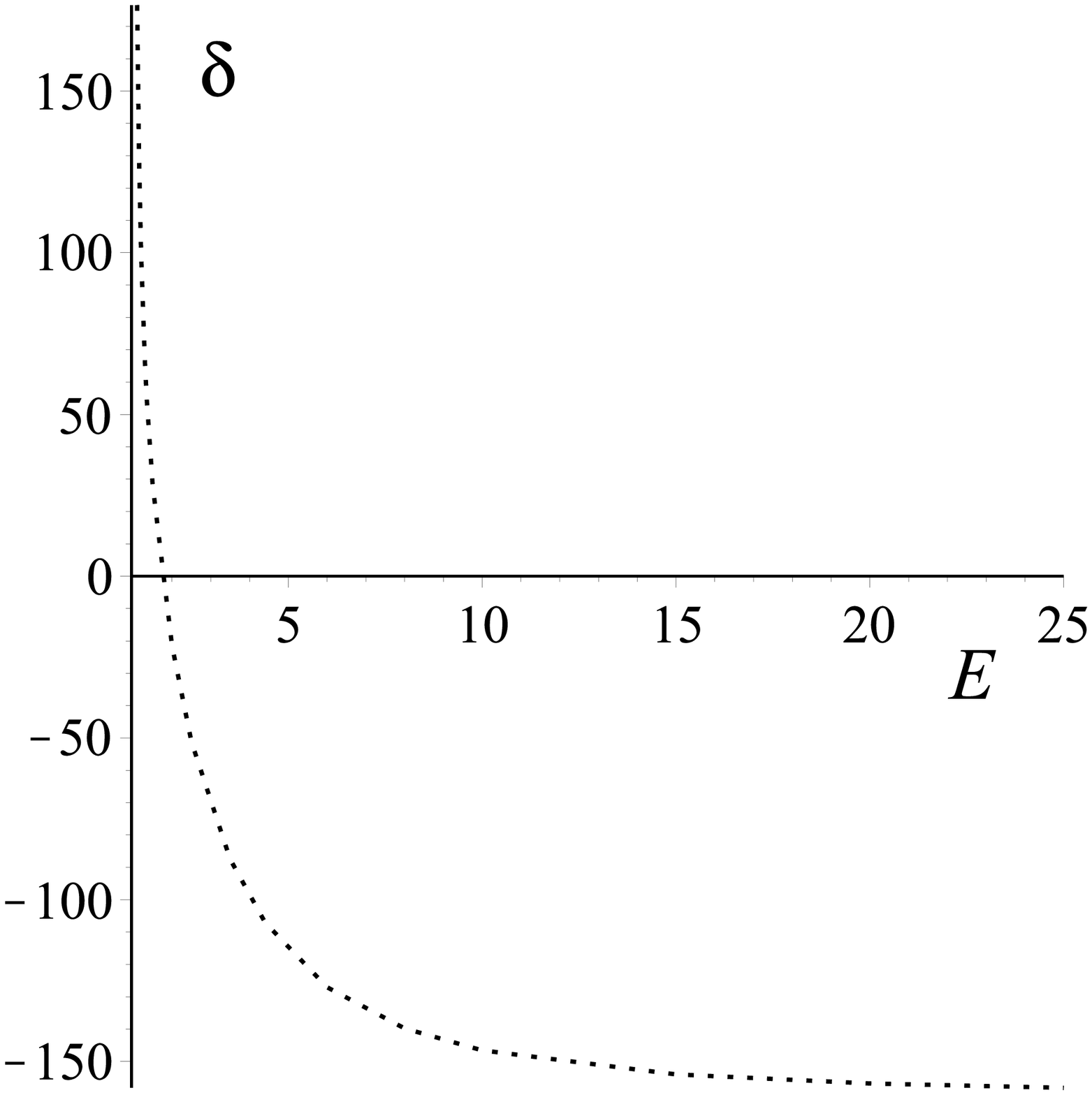}\\[.2cm]
\mbox{(a)} &\qquad \mbox{(b)}\cr
\end{array}
$\\
\end{center}
\caption{
Examples of numerical integration of scattering orbits in the $x-z$ plane are shown in panel (a) with the same parameter choice as in Fig. \ref{fig:orbite_asse_x} and initial conditions $x(0)=4$, $z(0)=-10$ and $\alpha(0)=\frac{\pi}{2}$ for every fixed value of energy.
The behavior of the scattering angle $\delta$ as a function of the energy is shown in panel (b).
For $M_2<M_1$ the curve moves to the left, since the energy for capture by the black hole with smaller mass $M_2$ (located in the same half plane ($z<0$) as the approaching trajectory) is less than in the equal mass case.
}
\label{fig:orbite_asse_z}
\end{figure}

Unbound orbits, analyzed mostly numerically due to the non-separability of the geodesic equations, are our main concern here.
We consider the distance $b$ between the two black holes as fixed, and study the features of the orbits in the $x-z$ plane (i.e., setting $y=0$, without loss of generality) by varying their energy parameter $E$.
The initial conditions are taken far from the system, with the particle moving towards it.
In the equal mass case there are two configurations summarizing the typical behavior of scattering orbits:
approach orthogonal to the $z-$axis (i.e., the line along which the black holes are displaced) and parallel to it. In both cases the orbits are either captured by one of the holes or undergo a scattering process, apart from the special orbit along the $x-$axis which proceeds undeflected for symmetry reasons.
The main gauge-invariant information uniquely characterizing the process is the scattering angle $\delta$, which can be evaluated numerically for each orbit looking at the asymptotic regime.
We have summarized the results of our analysis in Figs. \ref{fig:orbite_asse_x} and \ref{fig:orbite_asse_z}, where we have shown typical scattering orbits for different values of the energy as well as the associated scattering angle $\delta=\delta(E)$.
A peculiar feature of orbits approaching the system along a direction parallel to the $z-$axis is a \lq\lq S-like" trajectory, implying a change of sign in the scattering angle $\delta$ (see Fig. \ref{fig:orbite_asse_z}).

It can be useful to introduce a scattering matrix, which in this case turns out to be a rotation of the initial direction $\hat\nu$ of the spatial momentum in the $x-z$ plane, which can be written as
\beq
\pmatrix{
\hat\nu^{\hat x}_+& \hat\nu^{\hat z}_+
}
=
\pmatrix{
 \cos\delta&-\sin\delta\cr
 \sin\delta&\cos\delta
}
\pmatrix{ 
 \hat\nu^{\hat x}_-\cr
 \hat\nu^{\hat z}_-
}
\,,
\eeq
implying $\alpha_+-\alpha_-=2k\pi+\delta$ ($k$ integer), because of the conservative nature of the process considered.


\section{Gyroscope precession}

Let us consider a test gyroscope moving along a scattering orbit.
One can naturally define a frame adapted to $U_{\rm (geo)}$ by boosting the Cartesian-like frame \eqref{thdframe} along $U_{\rm (geo)}$ following a standard approach \cite{Jantzen:1992rg}, leading to 
\begin{eqnarray}
E(U_{\rm (geo)})_1&=&e_{\hat x}+\frac{\gamma\nu}{\gamma+1}(U_{\rm (geo)}+u)\cos\alpha\,,\nonumber\\
E(U_{\rm (geo)})_2&=&e_{\hat y}\,,\nonumber\\
E(U_{\rm (geo)})_3&=&e_{\hat z}+\frac{\gamma\nu}{\gamma+1}(U_{\rm (geo)}+u)\sin\alpha\,.
\end{eqnarray}
The transport properties of this frame along $U_{\rm (geo)}$ are 
\begin{eqnarray}
\label{fsframe}
\frac{DE(U_{\rm (geo)})_1}{d\tau}&=&\tau_1E(U_{\rm (geo)})_3\,,\nonumber\\
\frac{DE(U_{\rm (geo)})_2}{d\tau}&=&0\,,\nonumber\\
\frac{DE(U_{\rm (geo)})_3}{d\tau}&=&-\tau_1E(U_{\rm (geo)})_1\,,
\end{eqnarray}
where
\beq
\tau_1=-\frac{(2\gamma+1)(\gamma-1)}{2\gamma^2-1}\frac{d\alpha}{d\tau}\,, \qquad 
\gamma=\gamma(\tau)\,.
\eeq
The frame \eqref{fsframe} is a (degenerate, since the orbits are geodesic) Frenet-Serret frame along $U_{\rm (geo)}$, which can be cast in the familiar form by a reordering of the axes \cite{Iyer:1993qa}.
A further rotation of this frame along the $E(U_{\rm (geo)})_2$ axis with angular velocity $\tau_1$ then leads straightforwardly to a parallel transported frame along $U_{\rm (geo)}$. 

\begin{figure}
\begin{center}
$\begin{array}{cc}
\includegraphics[scale=0.25]{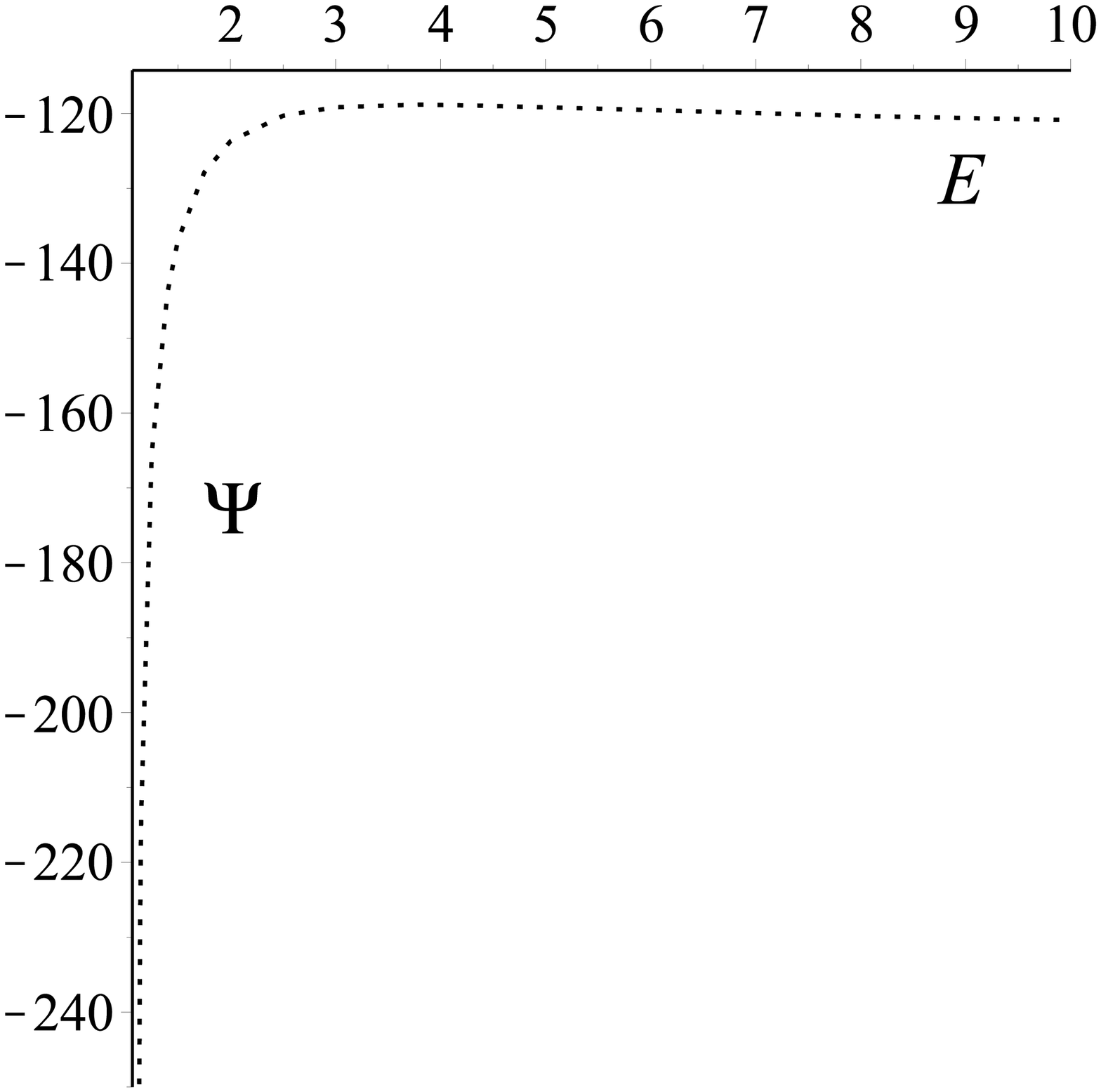}&\qquad
\includegraphics[scale=0.25]{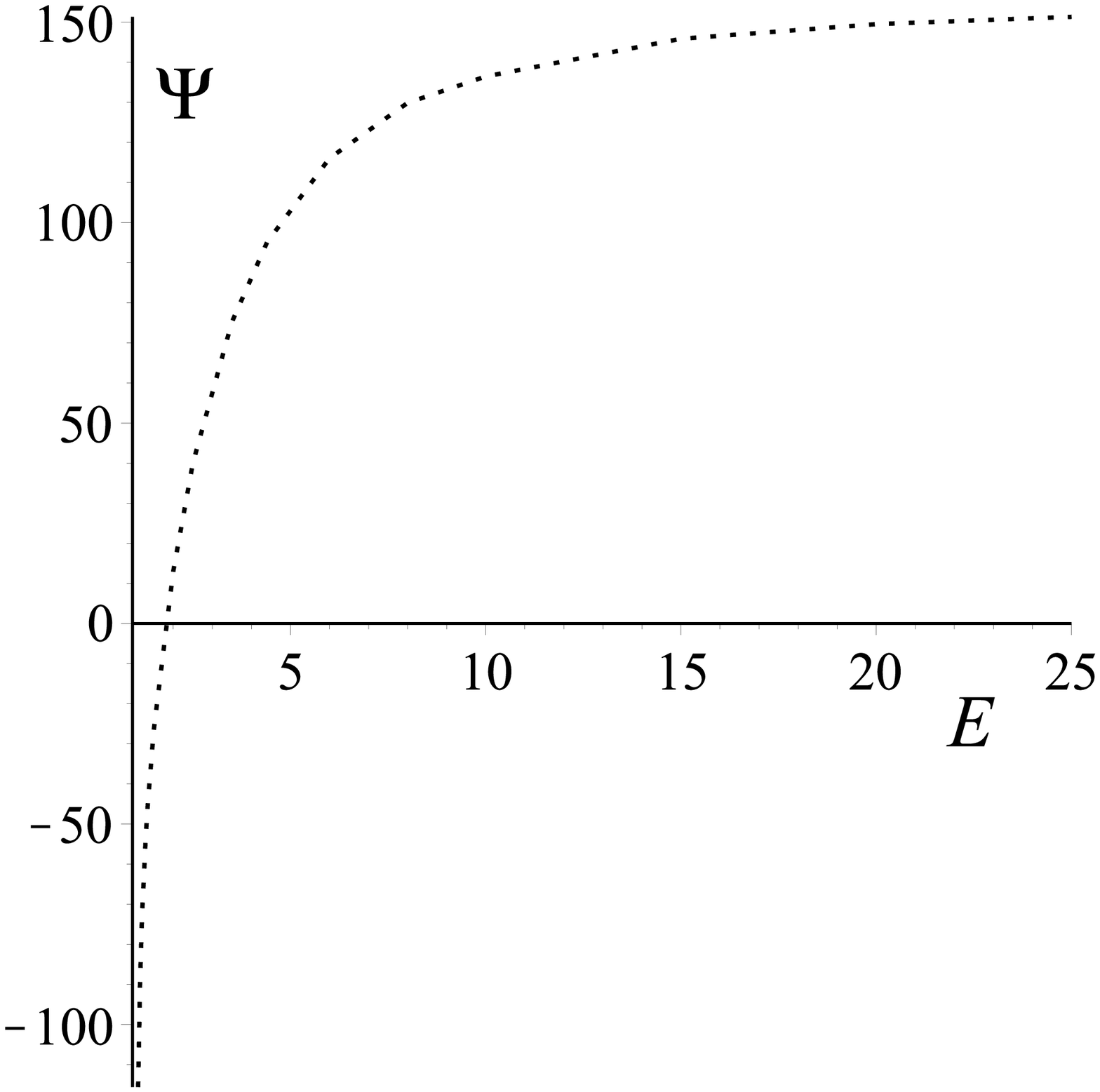}\\[.2cm]
\mbox{(a)} &\qquad \mbox{(b)}\cr
\end{array}
$\\
\end{center}
\caption{
The gyroscope precession angle $\Psi$ as a function of the energy for the orbits of Figs. \ref{fig:orbite_asse_x} and \ref{fig:orbite_asse_z} is shown in panels (a) and (b), respectively.
}
\label{fig:Psi}
\end{figure}

Finally, one can evaluate the total precession angle after a full scattering process by integrating $\tau_1$, i.e.,
\beq
\Psi=\int_{-\infty}^{\infty} \tau_1 d\tau\,,
\eeq
which is another gauge-invariant quantity often referred to as \lq\lq spin holonomy'' in the literature.
The behavior of $\Psi$ as a function of $E$ is shown in Fig. \ref{fig:Psi} for the same orbits as in Figs. \ref{fig:orbite_asse_x}--\ref{fig:orbite_asse_z}.
Again the change of sign in $\Psi$ is a specific marker of the \lq\lq S-like" orbits.


\section{Concluding remarks}

We have analyzed unbound timelike geodesic orbits of uncharged test particles moving in the (exact) gravitational and electromagnetic field of two extremely charged black holes described by the Majumdar-Papapetrou solution. Due to their \lq\lq extreme" property of having equal mass-to-charge ratios the two black holes are in a static equilibrium configuration irrespective of their distance. Despite the apparent simplicity of the metric, their mutual interaction causes complicated effects on the motion of massive particles as well as photons, in such a way that the geodesic equations (and the orbits of charged particles too) cannot be separated and exhibit chaotic behavior, even in the case of null orbits, which usually allows for further simplification.
This fact gives limitation to any analytical treatment, implying that the study of the dynamics can be performed mostly numerically.

Much attention in recent years has been devoted to bound motion of both particles and photons.
We have complemented this information here by analyzing the behavior of neutral (so to neglect electromagnetic interactions) particles undergoing a scattering process, i.e., starting their motion far from the system, approaching it up to a minimum approach distance, and then escaping back to spatial infinity.
We have (numerically) investigated the behavior of two different families of scattering orbits on a plane containing both black holes, corresponding to the two complementary situations of particles approaching the system along a direction parallel to the axis where the black holes are displaced, and orthogonal to it.
While one is familiar with typical hyperbolic-like orbits from a single body, in the present case of two scattering centers one should think instead in general of \lq\lq S-like" orbits, i.e., hyperbolic-like orbits with two different branches, mimicking attraction from one body in the first half of the motion and from the second body in the second half. 
Such a three-body scattering process (i.e., the two-body system and the scattered particle) represents a toy model for other possible sources of gravitational waves detectable by future advanced phase of Earth-based interferometers. 
In fact, the static background assumed here would be a dynamical spacetime in true processes, the two black holes undergoing head-on collision, so that it can only be a \lq\lq snapshot" of a real astrophysical collapse, namely an instantaneous configuration of a more complicated interaction picture. Nevertheless, even with all the limitations mentioned above, it is useful to have some hint or some expectation for this scenario.

We have summarized our results by computing two gauge-invariant quantities naturally associated with the scattering process: the scattering angle and the accumulated precession angle of a test gyroscope along such orbits.
The change of sign in both of them is a signature of S-like orbits in comparison with typical hyperbolic-like ones.
The extension of the present study to either charged or spinning particles also in the context of gravitational as well as electromagnetic perturbations of the background solution remains for future challenges.

\section*{Aknowledgments}

We thank Prof. O. Semer\'ak for useful discussions.

\end{document}